\documentclass[aps,prl,10pt,amsmath,floats,floatfix,twocolumn,superscriptaddress,altaffilletter,nofootinbib,shownopacs, frontmatterverbose,reprint]{revtex4-1}
\newcommand{\prlsec}[1]{%
  \par\vspace{2mm}
  \noindent\textbf{\textit{#1}}.---\,
  \ignorespaces
}

\usepackage{amsmath, amssymb}
\usepackage{listings}
\usepackage{graphicx}
\usepackage{dcolumn}
\usepackage{bm}
\usepackage{float,color}
\usepackage{xcolor}
\usepackage[normalem]{ulem}
\usepackage{tabularx}
\usepackage{mathtools} 
\usepackage{multirow}
\usepackage{amssymb}
\usepackage{acronym}
\usepackage{natbib}
\usepackage{scrextend}
\usepackage{bm}
\usepackage{graphicx}
\usepackage[colorlinks=true, allcolors=blue]{hyperref}

\newacro{bns}[BNS]{binary neutron star}
\newacro{ns}[NS]{Neutron Star}
\newacro{eos}[EoS]{equation of state}
\newacro{ce}[CE]{Cosmic Explorer}
\newacro{et}[ET]{Einstein Telescope}
\newacro{lvk}[LVK]{LIGO-Virgo-KAGRA}

\begin{document}

\title{
Probing Neutron Star Equation of State Universality with Gravitational Waves
}
\author{Praveer Tiwari}
\affiliation{Chennai Mathematical Institute, Siruseri, 603103, India}
\author{Ajit Kumar Mehta}
\affiliation{Chennai Mathematical Institute, Siruseri, 603103, India}
\affiliation{International Centre for Theoretical Sciences, Tata Institute of Fundamental Research, Bangalore 560089, India}
\author{K. G. Arun}
\affiliation{Chennai Mathematical Institute, Siruseri, 603103, India}
\affiliation{Max Planck Institute for Gravitational Physics (Albert Einstein Institute), D-30167 Hannover, Germany}
\affiliation{Leibniz Universit\"at Hannover, D-30167 Hannover, Germany}
\begin{abstract}
Current neutron star equation-of-state inferences assume that a common equation of state describes all neutron stars. However, {poorly understood dense-matter physics may give rise to distinct long-lived stellar subpopulations following different
mass--tidal-deformability relations, invalidating this assumption. In this \emph{Letter}, we demonstrate that next-generation gravitational-wave observatories enable neutron star equation-of-state universality to be transformed from an assumption into a testable hypothesis. By combining hierarchical Bayesian inference with a piecewise-polytropic parametrization of the neutron star equation of state, we establish a framework for testing equation-of-state universality with next-generation gravitational-wave observatories. We find that about 30 (50) loud binary neutron star mergers with network SNRs above 100 are sufficient to provide compelling evidence for departures from equation-of-state universality, even if only 20\% (10\%) of the population exhibits one of the representative non-universality parametrizations considered here.}

\end{abstract}

\maketitle

\prlsec{Introduction}
\acp{ns} are among the most intriguing astrophysical objects, offering a unique opportunity to study the interplay between strong gravity and ultra-dense matter. Their characteristic central densities can reach several times the nuclear saturation density \cite{Ozel2016, Baym2018}, making them unique astrophysical laboratories for probing the behavior of baryonic matter at extreme densities and pressures that are inaccessible to terrestrial experiments \cite{Lattimer2001, Lattimer2004, Lattimer2007, Ozel2016}.

Since neutron star cores contain matter at and above nuclear saturation density, and the underlying microphysics governing such matter is expected to be universal, current neutron star equation-of-state inference frameworks generally assume a universal equation of state~\cite{Hebeler2010, walker2024precision, Essick2020, Landry2019, Legred2021, Raaijmakers2021, pang2021nuclear, tsang2024determination, koehn2025existing, biswas2021impact, tiwari2023framework, huth2022constraining}. In other words, all observed NSs are assumed to share the same pressure-density relation. We refer to this as \ac{eos} universality in the remainder of this \emph{Letter.}

 While this assumption is well motivated, it remains a hypothesis about the NS population. The presence of exotic phases in neutron star cores does not, by itself, imply a violation of EoS universality, provided they are described by the same EoS for all neutron stars~\cite{LATTIMER2006479}. Given our limited understanding of matter at supra-nuclear densities, EoS universality should be tested rather than assumed. We therefore develop a framework to determine whether a single EoS can describe the observed neutron star population, without requiring knowledge of the microscopic origin of any departure.

Next-generation (XG) ground-based gravitational-wave observatories such as Cosmic Explorer~\cite{Reitze2019} and the Einstein Telescope~\cite{Punturo2010, Maggiore2020} are expected to detect and characterize tens of thousands of \acp{bns}, many with signal-to-noise ratios exceeding 100~\cite{Evans2021, colombo2025multi, hua2024population, muller2026distinguishing}. 
Measurements of the tidal deformabilities of this population will provide a unique opportunity to test the universality of the neutron star EoS. We demonstrate that a small subset of loud binary neutron star mergers in the XG era can place stringent constraints on, or detect, departures from EoS universality. To this end, we introduce a parametrization that quantifies such departures and apply it, within a hierarchical Bayesian framework, to a population of BNSs observable by next-generation gravitational-wave detectors.
\begin{figure*}[t]
    \includegraphics[width=0.48\textwidth]{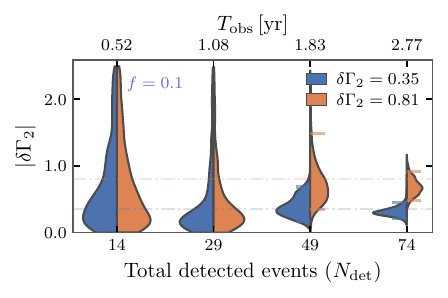}
    \includegraphics[width=0.48\textwidth]{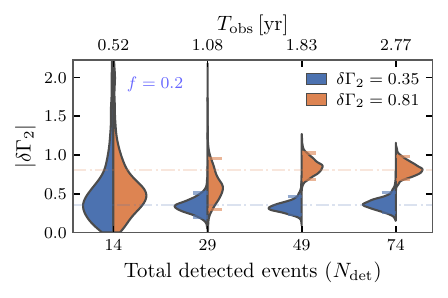}
    \caption{Posterior distributions for $|\delta\Gamma_2|$ as a function of the total number of detected events, $N_{\mathrm{det}}$. The absolute value is shown because the population likelihood is symmetric under $\delta\Gamma_2 \to -\delta\Gamma_2$. The upper x-axis indicates the corresponding observing time for the Cosmic Explorer--LIGO Hanford--LIGO Livingston network. The left and right panels show deformed mixing fractions $f=0.1$ and $f=0.2$, respectively, with split violins denoting injected populations at $\delta\Gamma_2=0.35$ and $0.81$. The posteriors tighten and move away from zero as $N_{\mathrm{det}}$ increases, with faster separation for larger $f$ and larger injected deviations. The dash-dotted lines show the corresponding injected values. Horizontal markers indicate the $90\%$ posterior probability intervals for cases in which zero lies outside the corresponding high-posterior-density region. For the smaller deviation, $\delta\Gamma_2=0.35$, distinguishing the deformed population requires $\sim 49$ loud events when $f$ is small.
}
    \label{fig:violin_plots}
\end{figure*}

\prlsec{Parametrization}
A piecewise-polytropic representation of the pressure-density relation is widely used as an effective phenomenological description of the NS EoS. In the pioneering work of Ref.~\cite{Read2009}, it was shown that a broad class of realistic EoSs can be accurately represented by a crust model combined with three polytropic segments that describe progressively higher-density regions of the NS interior. Within each segment, the pressure is assumed to follow a relation of the form $P \propto \rho^{\Gamma}$, and the segments are joined to obtain a continuous description of the NS EoS. 

We follow a similar prescription and adopt a fixed-crust model described by four polytropic segments and a core described by low-, intermediate-, and high-density regions. Due to the continuity conditions imposed at the interfaces between adjacent segments, the core EoS is characterized by four independent parameters, $\{\log p_1,\Gamma_1,\Gamma_2,\Gamma_3\}$, where $p_1$ denotes the pressure at a transition region between the low and the intermediate density regions and $\{\Gamma_1, \Gamma_2, \Gamma_3\}$ are the polytropic indices describing progressively higher-density regions of the NS core. Details can be found in the Supplementary Materials. We note that there are other parameterizations, both phenomenological~\cite{Raaijmakers2021, Lindblom2010} and physics conforming~\cite{forbes2019constraining, biswas2021impact, tiwari2023framework}, for the \ac{eos} of \ac{ns}. We choose a piecewise polytrope for computational efficiency, though the method can, in principle, be adapted to other classes of EoS parametrizations as well. 

 Gravitational-wave observations probe the neutron star EoS through measurements of tidal deformability. At densities below approximately $1.5\,n_0$, where $n_0$ is the nuclear saturation density, the EoS is relatively well constrained by nuclear experiments and many-body calculations, whereas at higher densities, astrophysical observations provide our only direct probes and significant uncertainties remain. Consequently, in our parametrized model, we keep the crust description fixed and parametrize departures from EoS universality through a deformation parameter, $\delta\Gamma_2$, applied to the polytropic index $\Gamma_2$ of the NS core. This choice targets the intermediate-density core (i.e., $1.89\,n_0 < n_B < 3.76\,n_0$), to which the tidal deformability is particularly sensitive, while keeping the model one-dimensional and computationally tractable.

Viewing this as a null parametrization, even if new physical effects such as metastable conversion states~\cite{Drago:2013fsa,Drago:2015cea} exist in the innermost regions of the NS, the parametrization would still be capable of detecting its presence, albeit suboptimally. {As a minimal null parametrization, the framework is agnostic about the physical origin of a departure from a unique population-wide mass--tidal relation. It may therefore capture the observable consequences of intrinsically distinct effective EoSs, star-dependent additional components, or multiple long-lived stellar families arising from common underlying microphysics, without explicitly modeling the corresponding microscopic mechanism.}

\begin{figure*}[t]
    \centering
    \includegraphics[width=0.48\textwidth]{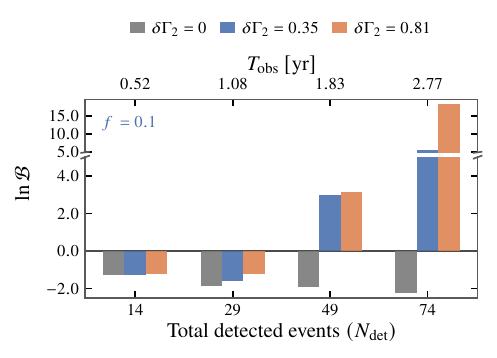}
    \includegraphics[width=0.48\textwidth]{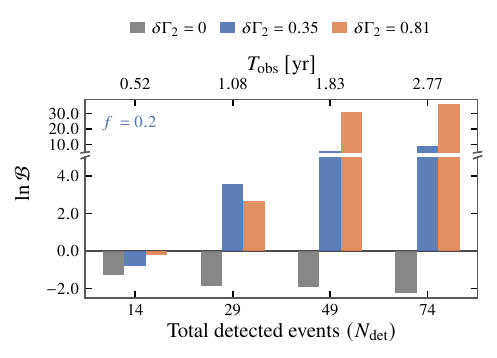}
    \caption{Log Bayes factor, $\ln \mathcal{B}$, comparing the mixed-population hypothesis to the baseline-population hypothesis for a deformed mixing fraction $f=0.1$ (left panel) and $f=0.2$ (right panel). The lower x-axis shows the number of detected events, $N_{\mathrm{det}}$, while the upper x-axis gives the corresponding optimistic observing time for the Cosmic Explorer--LIGO Hanford--LIGO Livingston network. The evidence for the mixed population becomes decisive when the posterior for $\delta\Gamma_2$ strongly excludes zero, as in Fig.~\ref{fig:violin_plots}.}
    \label{fig:bayes_factor}
\end{figure*}

\prlsec{Population inference framework} There are two specific questions that we seek to address. First, given a population of BNS mergers observed in the XG era, what departures from EoS universality can be detected as a function of the deformation amplitude $\delta\Gamma_2$ and the fraction of the population that violates universality? Second, if all NSs share a common EoS, how strongly can departures from universality be constrained as a function of the number of loud BNS detections?

 In our inference framework, at the individual-event level, we use a gravitational-wave waveform model that incorporates tidal effects (IMRPhenomXAS\_NRTidalv2~\cite{PhysRevD.102.064001, PhysRevD.100.044003, Colleoni:2023ple}) to infer the posterior distribution of the effective tidal deformability and other binary parameters. We then combine the information from multiple BNS events within a hierarchical Bayesian framework to infer the EoS parameters $\{
 %p_1,\Gamma_1,
\Gamma_2,\Gamma_3,\delta\Gamma_2\}$. Throughout this work, we adopt APR4~\cite{akmal1998equation} as the baseline EoS, with $\{\log p_1,\Gamma_1,\Gamma_2,\Gamma_3\}=\{34.269,\,2.830,\,3.445,\,3.348\}$~\cite{Read2009}.

To simulate departures from \ac{eos} universality, we vary the deformation parameter $\delta\Gamma_2$. We consider two representative deformation amplitudes, $\delta\Gamma_2=0.35$ and $0.81$, corresponding to small and moderate departures from universality, respectively. This enables us to construct synthetic source populations characterized by the deformation amplitude $\delta\Gamma_2$ and the fraction $f$ of systems that violate EoS universality. More specifically, we synthesize the following populations:
\begin{enumerate}
    \item All BNS systems follow the universal EoS ($\delta\Gamma_2=0$). This population allows us to assess how well the proposed method recovers consistency with EoS universality when it holds.
    \item Four populations corresponding to $\delta\Gamma_2=0.35,\,0.81$ and fractions $f=0.1,\,0.2$, which allow us to assess the ability of the method to detect non-universality as a function of the magnitude of the deviation and the fraction of non-universal systems in the population.
\end{enumerate}

\begin{figure*}[t]
    \centering
    \includegraphics[width=0.4\textwidth]{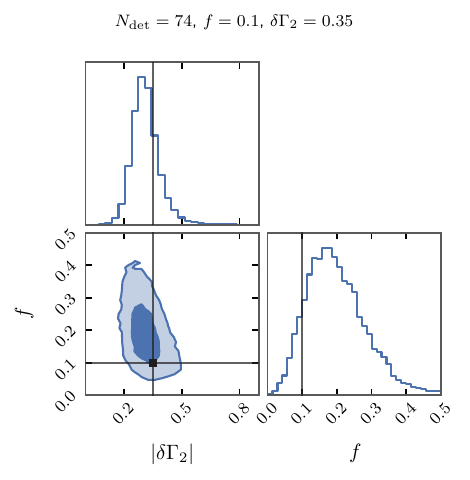}
    \includegraphics[width=0.4\textwidth]{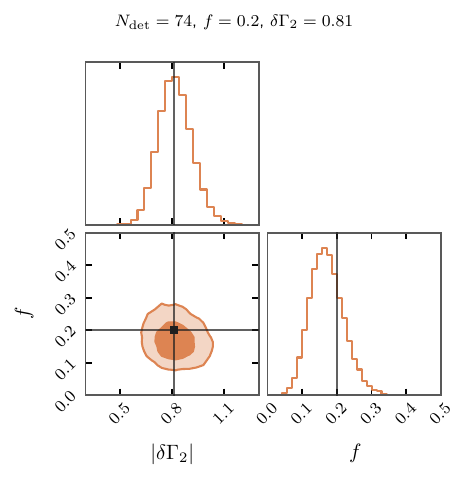}
    \caption{Corner plots of the deformation parameter $|\delta \Gamma_2|$ and the mixing fraction $f$ for two representative injected deviations, $\delta \Gamma_2=0.35$, $f=0.1$  (left panel) and $\delta \Gamma_2=0.81$, $f=0.2$ (right panel), with $N_{\mathrm{det}}=74$. The posterior samples have been folded under the label-switching symmetry, so the plotted fraction corresponds to $f_{\mathrm{mix}}=\min(f,1-f)$. The black reference lines indicate the corresponding injected values. The contours enclose $50\%$ and $90\%$ of the posterior probability. For the larger injected deviation, the posterior for the mixing fraction becomes more tightly constrained.
}
    \label{fig:corner_plots}
\end{figure*}

To simulate the BNS populations, we sample the component masses from a Gaussian distribution with mean $1.4\,M_{\odot}$ and standard deviation $0.68\,M_{\odot}$, truncated to the range $[1\,M_{\odot},M_{\rm max}]$, and impose $m_1\geq m_2$ \cite{ghosh2025joint}. Here, $M_{\rm max}$ is the maximum mass supported by the corresponding EoS.  We choose $1\,M_{\odot}$ as the lower mass bound, since current observations and population models suggest that BNS components below this mass are likely to be rare. This choice also keeps the component masses within the range for which the waveform models used for injection and recovery are designed and validated.

  We assume both neutron stars to be non-spinning, with their tidal deformabilities computed from the adopted piecewise-polytropic EoS, $\Lambda_i=\Lambda(m_i,\bm{\theta}_{\mathrm{EoS}})$, where $\bm{\theta}_{\mathrm{EoS}}$ denotes the parameters of the piecewise-polytropic EoS.  Consequently, the maximum mass of the baseline EoS model differs from the deformed one and this difference is consistently incorporated in the population synthesis. The luminosity distance is sampled uniformly in comoving volume, while the binary orientations are sampled isotropically by drawing the corresponding angular variables uniformly over the sphere. For simplicity and computational efficiency, the sky location is fixed to that of GW170817~\cite{Abbott2017GW170817, Abbott2018EoS, Abbott2019PRX}, which is expected to have little impact on our findings.

 We simulate $N$ BNS systems following the procedure described above, with $(1-f)N$ systems described by the baseline EoS ($\delta\Gamma_2=0$) and $fN$ systems described by the deformed EoS, where $f$ denotes the fraction of systems with $\delta\Gamma_2\neq0$ in the population. We then assess their detectability by computing their network SNRs for a detector network comprising LIGO Livingston~\cite{Aasi2015, KAGRA:2013rdx}, LIGO Hanford~\cite{Aasi2015, KAGRA:2013rdx}, and Cosmic Explorer~\cite{Reitze2019}, using the corresponding noise power spectral densities. We retain only those systems with network SNRs greater than 100, as these constitute the most informative events for our analysis. Bayesian inference is then performed on the loudest $N_{\rm det}$ events with SNRs above 100 in each population realization, where $N_{\rm det}$ is determined by the observing time of the detector network and estimated by computing the volume--time sensitivity for our baseline-EoS injection population (see the Supplemental Material for details).

 We consider $N_{\rm det}=14,\,29,\,49$, and $74$ to investigate how the sensitivity of the proposed method depends on the number of detected events. These values correspond approximately to 
 $[6.16, 0.2], [12.76, 0.42], [21.55, 0.71],$ and $[32.55, 1.07]$
 years of Cosmic Explorer operation, respectively, where the ranges reflect the current uncertainty in the local BNS merger rate,  $5.1$--$154.7~\mathrm{Gpc}^{-3}\,\mathrm{yr}^{-1}$~\cite{LIGOScientific:2026ctl}.

 We perform Bayesian parameter estimation on each of the $N_{\rm det}$ events in each population realization, using IMRPhenomXAS\_NRTidalv2~\cite{PhysRevD.102.064001, PhysRevD.100.044003, Colleoni:2023ple} as the waveform model and relative binning~\cite{krishna2023accelerated, Wong:2023lgb, Finstad:2020sok, Narola:2023men} to evaluate the likelihood. The likelihood is then sampled using the \texttt{dynesty} sampler~\cite{2020MNRAS.493.3132S} within the \texttt{Bilby} package~\cite{Ashton:2018jfp, Romero-Shaw:2020owr}. The priors on the source parameters follow the standard \ac{lvk} analysis conventions~\cite{LIGOScientific:2026ctl} and are summarized in Table~\ref{tab:pe_priors} of the Supplemental Material. Since we do not include spins in our analysis, the intrinsic BNS parameters relevant for our population analysis are the chirp mass $\mathcal{M}$, mass ratio $q$, and tidal deformabilities $(\Lambda_1,\Lambda_2)$.

We use the resulting event-level posterior samples to evaluate the population likelihood. The population hyperparameters include the EoS parameters $\Gamma_2$, $\Gamma_3$, and $\delta\Gamma_2$, together with the mixing fraction $f$; for computational efficiency, we fix $\log p_1$ and $\Gamma_1$. The hyperparameter set also includes the mean $\mu$ and standard deviation $\sigma$ of the truncated Gaussian distribution assumed for the population of component masses. The priors on the population hyperparameters used in this work are summarized in Table~\ref{tab:population_priors}. See the Supplemental Material for further details.

\prlsec{Results}
 We show our main results in Fig.~\ref{fig:violin_plots}. The figure shows the posterior distributions of the deformation parameter $|\delta\Gamma_2|$ for different numbers of detected events, $N_{\mathrm{det}}$, and different fractions of the deformed population, $f=0.1$ (left panel) and $f=0.2$ (right panel). We show only the absolute value of $\delta\Gamma_2$ because the population likelihood is symmetric under $\delta\Gamma_2\rightarrow-\delta\Gamma_2$ in our formulation. As expected, the posteriors become narrower and shift away from zero as $N_{\mathrm{det}}$ increases. When the fraction of deformed binaries is small and the deformation is also small ($f=0.1$ and $\delta\Gamma_2=0.35$), distinguishing the mixed population from the baseline population, i.e., $\delta\Gamma_2=0$, becomes feasible only for $N_{\rm det}\gtrsim49$. In contrast, for a larger fraction, $f=0.2$, and a larger deformation, $\delta\Gamma_2=0.81$, such a distinction becomes feasible for $N_{\rm det}\gtrsim29$. It is worth noting that the SNR distribution of the $N_{\rm det}$ selected events varies slightly between population realizations, depending on the number of events and other parameters governing the simulated population. Assuming a value around the median estimate of the BNS merger-rate density, i.e. $60~\mathrm{Gpc}^{-3}\,\mathrm{yr}^{-1}$ ~\cite{LIGOScientific:2026ctl}, our proposed method could become feasible within the first three years of observations with the Cosmic Explorer--LIGO Hanford--LIGO Livingston network~\cite{Reitze2019, Aasi2015, KAGRA:2013rdx}.

In Fig.~\ref{fig:bayes_factor}, we show the log Bayes factor, $\ln \mathcal{B}$, between the baseline and mixed-population hypotheses. The Bayes factor becomes appreciable, e.g., $\ln \mathcal{B}\gtrsim 2$, only when the posterior support for $|\delta\Gamma_2|$ is well separated from zero, as shown in Fig.~\ref{fig:violin_plots}. Otherwise, the Occam penalty associated with the two additional parameters, $\delta\Gamma_2$ and $f$, suppresses the evidence for the mixed-population model. For the smaller deformation, $\delta\Gamma_2=0.35$, and a small deformed fraction, $f=0.1$, the Bayes factor becomes positive and reaches $\ln \mathcal{B}\gtrsim 2$ only for $N_{\mathrm{det}}\geq 49$. This is also reflected in the left panel of Fig.~\ref{fig:corner_plots}, where the posterior for the fraction $f$ remains weakly constrained, even though the posterior for $|\delta\Gamma_2|$ begins to shift away from zero. In contrast, for the larger deformation, $\delta\Gamma_2=0.81$,  $f$ become well constrained and localized. As expected, larger deformations tend to produce larger Bayes factors, although this trend can be obscured at small $N_{\mathrm{det}}$ by finite-catalog fluctuations, when only a few events are drawn from the deformed subpopulation. As the number of detections increases, these realization-dependent fluctuations become subdominant and the expected trend with deformation strength becomes apparent. More generally, the Bayes factor increases as the posterior support for $|\delta\Gamma_2|$ moves farther from zero.

\begin{figure}[!htb]
    \includegraphics[width=\linewidth]{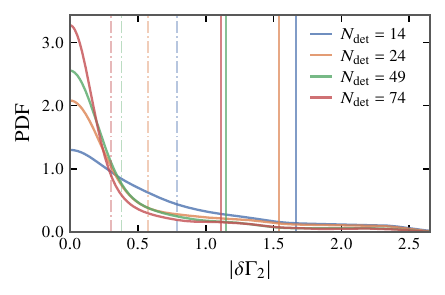}
    \caption{Posterior distributions for $|\delta\Gamma_2|$ under the baseline population for different $N_{\mathrm{det}}$. The dot-dashed vertical lines show the $68\%$ upper bounds while the solid lines show the 90\% upper bounds. Increasing $N_{\mathrm{det}}$ tightens the posterior around zero and improves the upper limit on deviations from the baseline EoS.
}
    \label{fig:pure_population}
\end{figure}

 Finally, if the EoS is universal, it is worth asking how strongly departures from universality can be constrained with a given number of loud detections. This provides a null test, complementary to the mixed-population searches discussed above. Instead of asking when a non-universal subpopulation can be identified, we ask how strongly event-to-event variations in the EoS can be constrained when the underlying population is described by a common EoS. In Fig.~\ref{fig:pure_population}, we show the posteriors for $|\delta\Gamma_2|$ and their 68\% and $90\%$ upper bounds, indicated by the vertical dot-dashed and solid lines respectively, as a function of $N_{\mathrm{det}}$. As expected, the posterior support shifts toward smaller values and the upper bound tightens as the number of loud detections increases. For $N_{\mathrm{det}}\sim74$, we find a $68\%$ upper bound of $|\delta\Gamma_2|\lesssim 0.3$, while the corresponding $90\%$ upper bound is $\lesssim 1.1$. The significant difference between the $68\%$ and $90\%$ upper bounds arises from the unexpected tail behaviour of the $|\delta\Gamma_2|$ posterior.

\begin{figure}[t]
    \centering
    \includegraphics[width=0.5\textwidth]{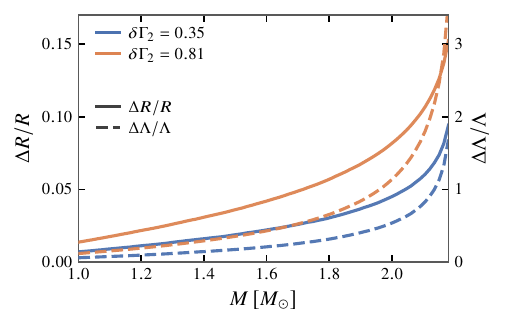}
    \caption{Fractional deviations in the radius and tidal deformability, relative to the baseline EoS, as functions of \ac{ns} mass for the two deformations considered in this work, $\delta\Gamma_2=0.35$ and $0.81$, corresponding to approximately $10\%$ and $24\%$ changes in the baseline value of $\Gamma_2$, respectively.}
    \label{fig:dG2RL}
\end{figure}

In the EoS parametrization used here, $\Gamma_2$ controls how rapidly the pressure increases with density over the intermediate-density core ($1.9\,n_0\lesssim n_B\lesssim3.76\,n_0$). The inferred bounds on $|\delta\Gamma_2|$ therefore constrain population-dependent variations in the EoS over this density range, rather than merely an abstract deformation parameter. Fig.~\ref{fig:dG2RL} maps the representative deformations considered in this work onto fractional differences in neutron star radii and tidal deformabilities. The figure underscores the important point that the departures accessible to our analysis do not require two dramatically different stellar configurations. For $\delta\Gamma_2=0.35$ ($0.81$), the radius difference remains below $5\%$ up to approximately $2.0\,M_\odot$ ($1.8\,M_\odot$). Nevertheless, the coherent accumulation of tidal information across the population can reveal these comparatively subtle differences, requiring $\sim49$ loud detections for $f=0.1$ and $\delta\Gamma_2=0.35$, and only $\sim29$ detections for $f=0.2$ and $\delta\Gamma_2=0.81$. Thus, a departure from EoS universality need not manifest as an individually exceptional neutron star; it may instead emerge statistically through a population of events whose tidal measurements are systematically better described by two distinct mass--tidal-deformability relations.

Our analysis makes a few simplifying assumptions. We neglect any modeling error arising from the finite flexibility of the phenomenological EoS parametrization. We use IMRPhenomXAS\_NRTidalv2~\cite{PhysRevD.102.064001, PhysRevD.100.044003, Colleoni:2023ple}, which models the adiabatic tidal response but neglects dynamical tidal effects. Finally, we assume the same mass-population model for the baseline and deformed subpopulations, which is sufficient for this proof-of-principle demonstration. We do not expect these simplifying assumptions to qualitatively affect the conclusions of our study, although they may modify the precise quantitative forecasts. Incorporating physically motivated EoS models, dynamical tides, and more general mass-population models will be important extensions of this framework.

To conclude, we have developed a Bayesian null test that turns the assumption of NS EoS universality into a population-level falsifiable hypothesis. We demonstrate that next-generation GW observatories, which will routinely observe BNS mergers, could carry out this test as early as within the first three years of operation, if we assume the current median estimate of the BNS merger-rate density. Whether they reveal departures from EoS universality or find the population consistent with a common EoS, the outcome will have profound implications for our understanding of dense matter and fundamental physics, highlighting the unique role of XG gravitational-wave observations as probes of matter under extreme conditions.

\section{Acknowledgements}
We thank Lami Suleiman for an internal review of this manuscript and several useful comments and suggestions. We thank Nathan Johnson- McDaniel, B. S. Sathyaprakash, Sanika S. Khadkikar, Debarati Chatterjee for useful discussions. P.T. and K.G.A. are supported by Advance Research Grant ANRF/ARG/2025/000931/PS of the Anusandhan National Research Foundation. K.G.A., P.T, and A.K.M. acknowledge support from the Infosys Foundation. K.G.A. acknowledges support from the Max Planck Society. P.T gratefully acknowledge the use of the high- performance super-computing cluster Kamiak at Washington State University (WSU).
This research has made use of data or software obtained from the Gravitational Wave Open Science Center (https://www.gwosc.org), a service of the LIGO Scientific Collaboration, the Virgo Collaboration, and KAGRA. This material is based upon work supported by NSF’s LIGO Laboratory which is a major facility fully funded by the National Science Foundation. This document has LIGO preprint number LIGO-P2600424.

\section{Supplemental material}
\prlsec{Piecewise-polytropic description of the neutron-star EoS}
\label{sec:pp_eos}

We model the low-density crust using a fixed four-piece polytropic
representation and parametrize the high-density EoS using three
polytropic segments,
\begin{equation}
    p(\rho)=K_i\rho^{\Gamma_i},
    \qquad
    \rho_{i-1}\leq \rho \leq \rho_i ,
\end{equation}
where $\rho$ is the rest-mass density. We closely follow the prescription
of Ref.~\cite{Read2009}, in which the high-density core of the NS is
divided into three regions with boundaries at
\begin{align}
    \rho_1 &= 10^{14.7}\ {\rm g\,cm^{-3}}\simeq 1.89\,n_0, \nonumber \\
    \rho_2 &= 10^{15.0}\ {\rm g\,cm^{-3}}\simeq 3.76\,n_0,
\end{align}
with $n_0=0.16~{\rm fm^{-3}}$. The high-density EoS is therefore
specified by
\begin{equation}
    \Upsilon=\{\log_{10}p_1,\Gamma_1,\Gamma_2,\Gamma_3\},
\end{equation}
where $p_1\equiv p(\rho_1)$. The polytropic constants are fixed by
requiring continuity of the pressure across the boundaries. In
particular,
\begin{equation}
    K_1=\frac{p_1}{\rho_1^{\Gamma_1}},
    \qquad
    K_2=\frac{p_1}{\rho_1^{\Gamma_2}},
\end{equation}
and
\begin{equation}
    K_3
    =
    \frac{p_1(\rho_2/\rho_1)^{\Gamma_2}}
         {\rho_2^{\Gamma_3}}.
\end{equation}
The lowest-density core segment is extended downward until it
intersects the fixed crust EoS. If the final crust segment is written
as $p=K_{c,4}\rho^{\Gamma_{c,4}}$, the joining density is
\begin{equation}
    \rho_0=
    \left(\frac{K_1}{K_{c,4}}\right)^{
    1/(\Gamma_{c,4}-\Gamma_1)} .
\end{equation}

In units with $c=1$, thermodynamic consistency gives the energy
density in segment $i$ as
\begin{equation}
    \epsilon(\rho)
    =
    (1+a_i)\rho+
    \frac{K_i}{\Gamma_i-1}\rho^{\Gamma_i},
\end{equation}
where the constants $a_i$ are chosen such that $\epsilon(\rho)$ is
continuous across every transition, including the crust--core
transition. This ensures thermodynamic consistency across the
piecewise-polytropic boundaries. \citet{suleiman2021influence,
suleiman2022polytropic} have performed a detailed analysis of the
errors introduced when mapping piecewise-polytropic EoS models onto
named EoS curves; we do not consider these modeling uncertainties
here.

For each EoS, we solve the Tolman--Oppenheimer--Volkoff and
quadrupolar tidal-perturbation equations to obtain the stable
mass--radius and mass--tidal-deformability relations. The dimensionless
tidal deformability is
\begin{equation}
    \Lambda(M)=\frac{2}{3}k_2(M)
    \left[\frac{R(M)}{M}\right]^5 ,
\end{equation}
where $k_2$ is the quadrupolar Love number.

Our objective is a minimal population-level null test rather than a
detailed model of any particular microscopic mechanism. We therefore
allow the two BNS subpopulations to differ only in $\Gamma_2$, while
keeping $\log_{10}p_1$, $\Gamma_1$, and $\Gamma_3$ the same for both.
The $\Gamma_2$ segment covers densities from approximately
$1.89\,n_0$ to $3.76\,n_0$ and spans a larger fraction of the density
range sampled by the BNS mass distribution than the $\Gamma_3$ segment.
Varying $\Gamma_2$ consequently produces observable differences over
a broader mass range while retaining a one-dimensional deformation.
The parameter $\delta\Gamma_2$ should therefore be interpreted as a
phenomenological measure of separation between two effective
mass--tidal-deformability relations, rather than as a direct
microphysical phase-transition parameter.

\prlsec{Hierarchical Bayesian inference and selection effects}
\label{sec:hbi}

We denote the mass-population hyperparameters by $\bm{\lambda}$ and
the EoS hyperparameters by $\bm{\Upsilon}$.  Let
\begin{equation}
    \tilde{\theta}
    \equiv
    \{\mathcal{M},q,\tilde{\Lambda}\}
\end{equation}
denote the intrinsic parameters entering the population analysis, and
let $\theta=(\tilde{\theta},z)$ additionally include redshift.  Here,
$\mathcal{M}$ is the source-frame chirp mass, $q=m_2/m_1\leq 1$, and
\begin{align}
    \tilde{\Lambda}
    =
    \frac{16}{13}
    \frac{
    (m_1+12m_2)m_1^4\Lambda_1+
    (m_2+12m_1)m_2^4\Lambda_2
    }{(m_1+m_2)^5}
\end{align}
is the effective tidal deformability { where $\Lambda_i$ is the dimensionless tidal deformability of the $i$-th component of \ac{bns}}.

The intrinsic population distribution
$p_{\rm int}(\tilde{\theta}\mid\bm{\lambda},\bm{\Upsilon})$
is normalized for every choice of the hyperparameters.  For the
redshift dependence, however, it is convenient to use the
unnormalized observer-frame source-count measure
\begin{equation}
    \mathcal{Q}(\theta\mid\bm{\lambda},\bm{\Upsilon})
    =
    p_{\rm int}(\tilde{\theta}\mid\bm{\lambda},\bm{\Upsilon})
    \frac{\psi(z)}{1+z}\frac{dV_c}{dz},
    \label{eq:unnormalized_population_measure}
\end{equation}
where $\mathcal{R}(z)=\mathcal{R}_0\psi(z)$ is the source-frame
merger-rate density and $\psi(0)=1$.  We set $\psi(z)=1$ throughout
this work.  We reserve $f$ for the mixing fraction of the nonstandard
EoS subpopulation.

For $N_{\rm det}$ independent detected events with data
$\{d_i\}$, the inhomogeneous-Poisson likelihood can be written \cite{Roulet:2020wyq, Mehta:2025oge},
up to factors independent of the hyperparameters, as 
\begin{widetext}
\begin{align}
    p(\{d_i\},N_{\rm det}\mid
    \bm{\lambda},\bm{\Upsilon},\mathcal{R}_0)
    \propto
    \mathcal{R}_0^{N_{\rm det}}
    e^{-\mathcal{R}_0\overline{VT}(
    \bm{\lambda},\bm{\Upsilon})}
    \prod_{i=1}^{N_{\rm det}}
    \int d\theta\,
    \mathcal{L}(d_i\mid\theta)
    \mathcal{Q}(\theta\mid
    \bm{\lambda},\bm{\Upsilon}) .
    \label{eq:full_poisson_likelihood}
\end{align}
\end{widetext}
Fixed factors of the observing time in the per-event terms have been
absorbed into the overall normalization.  The population-averaged
sensitive spacetime volume is
\begin{equation}
    \overline{VT}(\bm{\lambda},\bm{\Upsilon})
    =
    T_{\rm obs}
    \int d\theta\,
    p_{\rm det}(\theta)
    \mathcal{Q}(\theta\mid
    \bm{\lambda},\bm{\Upsilon}),
    \label{eq:VT}
\end{equation}
where $p_{\rm det}(\theta)$ is averaged over the extrinsic parameters
not included explicitly in $\theta$.

We do not infer the absolute merger-rate density.  Marginalizing
$\mathcal{R}_0$ with the scale-invariant prior
$\pi(\mathcal{R}_0)\propto 1/\mathcal{R}_0$ gives
\begin{widetext}
\begin{align}
    p(\{d_i\},N_{\rm det}\mid
    \bm{\lambda},\bm{\Upsilon})
    \propto
    \prod_{i=1}^{N_{\rm det}}
    \frac{
    \displaystyle
    \int d\theta\,
    \mathcal{L}(d_i\mid\theta)
    \mathcal{Q}(\theta\mid
    \bm{\lambda},\bm{\Upsilon})
    }{
    \overline{VT}(\bm{\lambda},\bm{\Upsilon})
    } .
    \label{eq:rate_marginalized_likelihood}
\end{align}
\end{widetext}
This is equivalent to the conventional formulation written in terms
of a normalized redshift distribution and the dimensionless
selection efficiency $\xi$.  
The normalization of the redshift
measure cancels between the event terms and
$\overline{VT}^{\,N_{\rm det}}$ because the same unnormalized measure
in Eq.~\eqref{eq:unnormalized_population_measure} is used in both.

For event $i$, posterior samples obtained using the parameter-estimation
prior $\pi_{\rm PE}(\theta)$ satisfy
\begin{equation}
    p(\theta\mid d_i)
    =
    \frac{\mathcal{L}(d_i\mid\theta)
    \pi_{\rm PE}(\theta)}{\mathcal{Z}_i}.
\end{equation}
The event contribution can therefore be evaluated by posterior
reweighting,
\begin{equation}
    \int d\theta\,
    \mathcal{L}(d_i\mid\theta)\mathcal{Q}(\theta\mid\cdots)
    =
    \mathcal{Z}_i
    \int d\theta\,
    p(\theta\mid d_i)
    \frac{\mathcal{Q}(\theta\mid\cdots)}
         {\pi_{\rm PE}(\theta)}.
\end{equation}
The event evidence $\mathcal{Z}_i$ is independent of the population
hyperparameters and is omitted from the population likelihood.

In our implementation, redshift and the remaining extrinsic
parameters are marginalized at the event level, and the posterior
samples are transformed consistently to source-frame masses using
the {Planck15} cosmology~\cite{ade2016planck}.  The reduced event term is then
\begin{widetext}
\begin{align}
    \mathcal{L}_{i,\rm pop}
    \propto
    \frac{1}{\overline{VT}(\bm{\lambda},\bm{\Upsilon})}
    \int d\mathcal{M}\,dq\,d\tilde{\Lambda}\,
    p(\mathcal{M},q,\tilde{\Lambda}\mid d_i)
    \frac{
    p_{\rm int}(\mathcal{M},q,\tilde{\Lambda}
    \mid\bm{\lambda},\bm{\Upsilon})
    }{
    \pi_{\rm PE}(\mathcal{M},q,\tilde{\Lambda})
    }.
    \label{eq:posterior_reweighting}
\end{align}
\end{widetext}

\prlsec{The BNS subpopulation model}
\label{sec:population_model}

We model the BNS population as a mixture of two subpopulations that
follow distinct effective EoSs,
\begin{align}
    p_{\rm int}(&\tilde{\Lambda},\mathcal{M},q
    \mid\bm{\lambda},\bm{\Upsilon})
    =
    (1-f)\,
    p_1(\tilde{\Lambda},\mathcal{M},q
    \mid\bm{\lambda},\Upsilon_1)
    \nonumber\\
    &\hspace{2.7cm}
    +f\,
    p_2(\tilde{\Lambda},\mathcal{M},q
    \mid\bm{\lambda},\Upsilon_2),
    \label{eq:mixture_population}
\end{align}
where $f$ is the fraction assigned to the second subpopulation.  For
each EoS branch,
\begin{align}
    p_a(&\tilde{\Lambda},\mathcal{M},q
    \mid\bm{\lambda},\Upsilon_a)
    =
    p_m(\mathcal{M},q\mid\bm{\lambda},\Upsilon_a)
    \nonumber\\
    &\quad\times
    \delta\!\left[
    \tilde{\Lambda}
    -\tilde{\Lambda}(\mathcal{M},q;\Upsilon_a)
    \right],
    \qquad a\in\{1,2\}.
    \label{eq:eos_branch_population}
\end{align}
The deterministic function
$\tilde{\Lambda}(\mathcal{M},q;\Upsilon_a)$ is obtained from the
mass--tidal-deformability relation generated by $\Upsilon_a$.

We define
\begin{equation}
    \Upsilon_1
    =
    \{\log_{10}p_1,\Gamma_1,\Gamma_2,\Gamma_3\},
\end{equation}
and adopt the convention
\begin{equation}
    \Upsilon_2
    =
    \{\log_{10}p_1,\Gamma_1,
      \Gamma_2+\delta\Gamma_2,\Gamma_3\}.
    \label{eq:second_eos_definition}
\end{equation}
Thus $\delta\Gamma_2=0$ corresponds to our baseline EoS model.  If the
implementation uses the opposite sign convention, the definition in
Eq.~\eqref{eq:second_eos_definition} should be reversed consistently
throughout the Letter and Supplemental Material.

For branch $a$,  following \cite{ghosh2025joint}, the component masses are generated independently
from a Gaussian with mean $\mu$ and standard deviation $\sigma$,
truncated to
\begin{equation}
    m_{\min}\leq m\leq M_{\max}(\Upsilon_a),
    \qquad m_{\min}=1\,M_\odot,
\end{equation}
and subsequently ordered so that $m_1\geq m_2$.  If
$g_a(m\mid\mu,\sigma)$ denotes the normalized truncated Gaussian,
the ordered joint distribution is
\begin{align}
    p_m(m_1,m_2\mid\mu,\sigma,\Upsilon_a)
    =&\,
    2\,g_a(m_1\mid\mu,\sigma)
      g_a(m_2\mid\mu,\sigma) \times \nonumber \\
      &\Theta(m_1-m_2).
\end{align}
The corresponding distribution in $(\mathcal{M},q)$ includes the
Jacobian of the transformation from $(m_1,m_2)$.  Although the same
$\mu$ and $\sigma$ are shared by the two subpopulations, their
normalized mass distributions can differ through the
EoS-dependent upper limit $M_{\max}(\Upsilon_a)$.

We fix $\log_{10}p_1$ and $\Gamma_1$ and infer
\begin{equation}
    \bm{\eta}
    \equiv
    \{\mu,\sigma,\Gamma_2,\Gamma_3,
      \delta\Gamma_2,f\}.
    \label{eq:hyperparameter_set}
\end{equation}

The mixture is invariant under relabeling the two branches,
\begin{equation}
    (\Gamma_2,\delta\Gamma_2,f)
    \longleftrightarrow
    (\Gamma_2+\delta\Gamma_2,
     -\delta\Gamma_2,1-f),
    \label{eq:label_symmetry}
\end{equation}
provided the hyperpriors respect the same symmetry.  This produces
the label-switching degeneracy discussed in the Letter.  We therefore
display the deformation using $|\delta\Gamma_2|$ and, when a
label-independent fraction is required, use
\begin{equation}
    f_{\rm mix}=\min(f,1-f).
\end{equation}

\prlsec{ Synthesized population and selection-function calculation}
\label{sec:injections}

The source-frame differential merger rate is written as
\begin{equation}
    \frac{d^3N}
    {dt_s\,dV_c\,d\tilde{\theta}}
    =
    \mathcal{R}_0\,
    \psi(z)\,
    p_{\rm int}(\tilde{\theta}
    \mid\bm{\lambda},\bm{\Upsilon}).
\end{equation}
The corresponding expected number of detections is
\begin{equation}
    N_{\rm exp}
    =
    \mathcal{R}_0\,
    \overline{VT}(\bm{\lambda},\bm{\Upsilon}),
\end{equation}
with $\overline{VT}$ given by Eq.~\eqref{eq:VT}.

We approximate the detection probability by a hard threshold on the
network optimal SNR,
\begin{equation}
    p_{\rm det}(\theta)
    =
    \Theta[\rho_{\rm net}(\theta)-\rho_\ast],
    \qquad
    \rho_\ast=100.
\end{equation}
We choose this relatively high threshold to focus on the loudest
(``golden'') BNS events, which are expected to provide the most
informative measurements of the tidal effects considered in this work.

We evaluate $\overline{VT}$ using importance sampling.  Let
$\theta_j$ be $N_{\rm inj}$ simulated systems drawn from a normalized
reference distribution $p_{\rm ref}(\theta)$.  Then
\begin{equation}
    \overline{VT}(\bm{\lambda},\bm{\Upsilon})
    \simeq
    \frac{T_{\rm obs}}{N_{\rm inj}}
    \sum_{j=1}^{N_{\rm inj}}
    \mathbb{I}_j
    \frac{
    \mathcal{Q}(\theta_j
    \mid\bm{\lambda},\bm{\Upsilon})
    }{
    p_{\rm ref}(\theta_j)
    },
    \label{eq:vt_monte_carlo}
\end{equation}
where
$\mathbb{I}_j=\Theta[\rho_{{\rm net},j}-\rho_\ast]$.  Equivalently,
the sum may be restricted to the found injections, in which case
$\mathbb{I}_j$ is omitted.  We choose our reference distribution such that it covers
the support of every population model allowed by the hyperpriors.

To generate an astrophysical catalog, we first assign each binary to
the standard or nonstandard branch with probabilities $1-f$ and $f$.
We then draw the component masses from the branch-dependent truncated
Gaussian, solve for $\Lambda_1$ and $\Lambda_2$ using the corresponding
EoS, and calculate $\tilde{\Lambda}$.  For a constant source-frame
merger-rate density, the redshift distribution of events in detector
time is proportional to
\begin{equation}
    p(z)\propto
    \frac{1}{1+z}\frac{dV_c}{dz}.
    \label{eq:catalog_redshift_distribution}
\end{equation}

The sky position is fixed to the GW170817 values, and
the remaining extrinsic parameters are drawn from the distributions
listed in Table~\ref{tab:pe_priors}. The masses for the \ac{bns} population is sampled from truncated Gaussian (between $1 M_{\odot}$ and $M_{\rm max}$) with mean $1.4 M_{\odot}$ and standard deviation $0.68 M_{\odot}$, ensuring $m_1 \geq m_2$.

\begin{figure*}
    \centering
    \includegraphics[width=\textwidth]{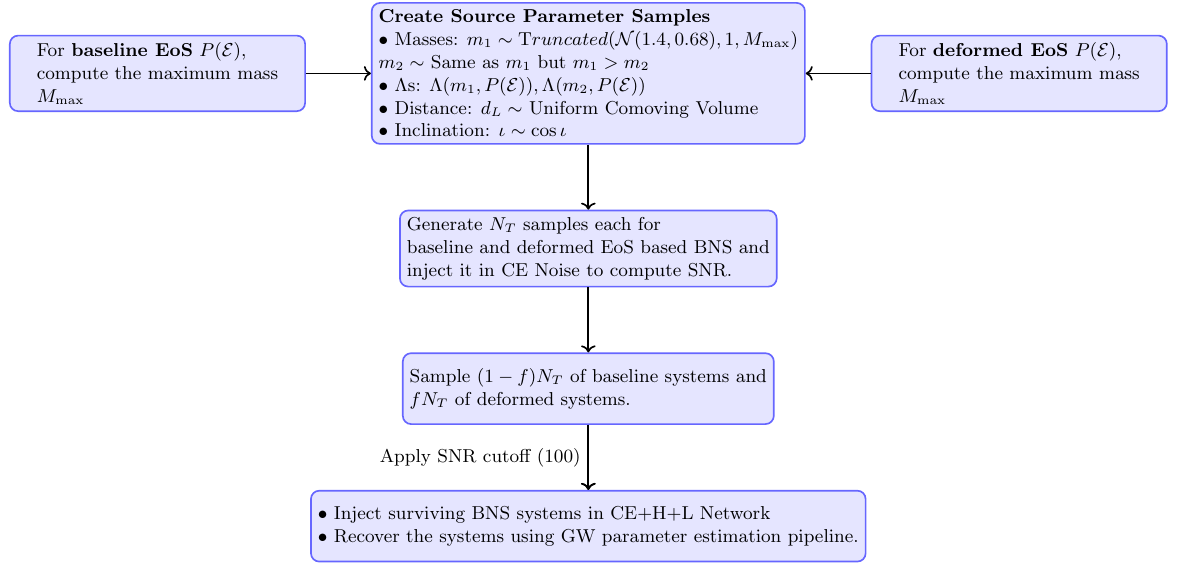}
    \caption{
    Flowchart summarizing the generation, selection, event-level
    parameter estimation, and hierarchical analysis of the simulated
    BNS catalogs.
    }
    \label{fig:inj_flow}
\end{figure*}

For every simulated binary, we calculate the network SNR for the
Cosmic Explorer--LIGO Hanford--LIGO Livingston network and retain
systems with $\rho_{\rm net}>100$.  The retained signals are injected
into independent Gaussian-noise realizations generated using the
detector power spectral densities adopted in the Letter.  We then
perform event-level parameter estimation and retain posterior samples
for the intrinsic parameters used in
Eq.~\eqref{eq:posterior_reweighting}.  A flowchart of the full procedure is
shown in Fig.~\ref{fig:inj_flow}.

\prlsec{Priors and injected hyperparameters}
\label{sec:priors}

Tables~\ref{tab:pe_priors} and~\ref{tab:population_priors} list all the priors used in the event-level and population-level analyses. We note that the prior ranges of the EoS parameters contain points that violate causality. However, given that the SNR and the number of detections are sufficiently high, the likelihood at noncausal points becomes infinitesimal. This can be seen from the posteriors of the recovered EoS parameters.

\begin{table*}
    \centering
    \caption{
    Priors used for event-level parameter estimation.
    }
    \label{tab:pe_priors}
    \begin{tabular}{llll}
        \hline\hline
        Parameter & Prior & Range & Comment\\
        \hline
        Chirp mass & Uniform & [0.4, 4.4] $M_{\odot}$ & Detector frame\\
        Mass ratio $q$ & Uniform & [0.125, 1] & $q\leq1$\\
        $\Lambda_1,\Lambda_2$ & Uniform & [0, 5000] & \\
        Luminosity distance & Uniform & [100, 5000] Mpc & Uniform in comoving volume and source-frame time (Planck15)\\
        Inclination & Sine & [0, $\pi$] & --\\
        Right Ascension & fixed & 3.44616 & --\\
        Declination & fixed & -0.40808 & --\\
        Polarization & fixed & 2.659 & --\\
        Coalescence phase & fixed & 1.3 & \\
        Spins & fixed & $0.02254$ & Magnitudes\\
        \hline\hline
    \end{tabular}
\end{table*}

\begin{table*}
    \centering
    \caption{
    Hyperpriors and injected values used in the hierarchical analysis.
    }
    \label{tab:population_priors}
    \begin{tabular}{llll}
        \hline\hline
        Hyperparameter & Hyperprior & Injection(s) & Comment\\
        \hline
        $\mu$ & Uniform [1.2, 1.6] & 1.4 & Mass-distribution mean\\
        $\sigma$ & Uniform [0.5, 0.9] & 0.68 & Mass-distribution width\\
        $\log_{10}p_1$ & fixed & $34.269$ & $\log_{10} \rm (dyne/cm^2)$\\
        $\Gamma_1$ & fixed & $2.830$ & \\
        $\Gamma_2$ & Uniform [1, 6] & $3.445$ & Baseline branch\\
        $\Gamma_3$ & Uniform [2, 6] & $3.348$ & Shared by both branches\\
        $\delta\Gamma_2$ & Uniform [-2.5, 2.5] & $0,\ 0.35,\ 0.81$ & \\
        $f$ & Uniform [0, 1] & $0.1,\ 0.2$ & \\
        \hline\hline
    \end{tabular}
\end{table*}

\bibliography{sample}
\end{document}